\documentclass[aps,superscriptaddress,showpacs,amsmath,amssymb,amsfonts,altaffillsymbol,twocolumn,notitlepage,prl]{revtex4-1}

\usepackage{graphicx}% Include figure files
\usepackage{float}
\usepackage{dcolumn}% Align table columns on decimal point
\usepackage{bm}
\usepackage{hyperref}

\graphicspath{{./figures/}}

\begin{document}
\title{Networks and Hierarchies: How Amorphous Materials Learn to Remember}

\date{\today}% It is always \today, today,

\author{Muhittin Mungan}
\email[Corresponding author: ]{mungan@iam.uni-bonn.de}
\affiliation{Institut f\"{u}r angewandte Mathematik, Universit\"{a}t Bonn, Endenicher Allee 60, 53115 Bonn, Germany}

\author{Srikanth Sastry}
\affiliation{Jawaharlal Nehru Centre for Advanced Scientific Research, Jakkar Campus, 560064 Bengaluru, India}

\author{Karin Dahmen}
\affiliation{Department of Physics, University of Illinois at Urbana-Champaign, 1110 West Green Street, Urbana, IL 61801, USA}

\author{Ido Regev}
\email[Corresponding author: ]{regevid@bgu.ac.il}
\affiliation{Jacob Blaustein Institutes for Desert Research, Ben-Gurion University of the Negev, Sede Boqer Campus 84990, Israel}

\begin{abstract}
%MM: slightly shorter abstract
We consider the slow and athermal deformations of amorphous solids and show how the ensuing sequence of discrete plastic rearrangements can be mapped onto a directed network. The network topology reveals a set of highly connected regions joined by occasional one-way transitions. The highly connected regions include hierarchically organized hysteresis cycles and sub-cycles. At small to moderate strains this organization leads to near-perfect return point memory. The transitions in the network can be traced back to localized particle rearrangements (soft-spots) that interact via Eshelby-type deformation fields. By linking topology to dynamics, the network representations provides new insights into the mechanisms that lead to reversible and irreversible behavior in amorphous solids.
\end{abstract}

\maketitle

A wide variety of condensed matter systems exhibit memory effects, since their present states are a result of their past history, which is encoded in their structure. Often all or at least part of such histories  may be inferred from measurements \cite{keimetalRMP2019}. Examples include 
shape memory materials, disordered magnets, spin glasses, structural glasses and granular matter, and magnetic and phase change memory devices.  
In particular, memory effects in cyclically driven (sheared) amorphous solids and colloidal suspensions have been recently investigated through computer simulations, experiments and theoretical modeling \cite{keimetalRMP2019,Keimetal2014,fiocco2014encoding,Fiocco2015,munganwitten2018}. 
For small to moderate deformations, upon repeated cyclic loading, after a transient, these systems  reach limit cycles in which they traverse the same sequence of states during each subsequent cycle  \cite{corte2008random,Keimetal2011,Keimetal2013,regev2013onset,fiocco2013oscillatory,Keimetal2014,keim2014mechanical,fiocco2014encoding,regev2015reversibility,Fiocco2015,nagamanasa2014experimental,kawasaki2016macroscopic,leishangthem2017yielding,adhikari2018memory,keim2018return,keimetalRMP2019,regev2018critical,bandi2018training,libal2012hysteresis,munganwitten2018}. 
%The nature of memory effects seen in these systems are closely linked to the presence of these limit cycles. 

In contrast, systems obeying the “no-passing” property, an ordering of states that is preserved by the dynamics, 
%\footnote{A partial order is an order relation defined only for a subset of elements.}, 
exhibit limit cycles immediately, {\it i.e.} without any transients.
Examples include systems with either no coupling at all, such as the Preisach model \cite{Preisach1935} or systems that have only positive couplings, such as depinning models and the random field Ising model \cite{middleton1992asymptotic}.   
Theoretical studies show that “no-passing” is a sufficient condition for return point memory (RPM) \cite{middleton1992asymptotic,sethna1993hysteresis}, wherein a system remembers the values at which the direction of an external driving field are reversed. Negative couplings can break the no passing property. 
% MM: This sentence is too long
%Indeed, in amorphous solids units of plastic deformation lead to state changes (described further below) that are referred to as shear transformation zones \cite{argon1979plastic,falk1998dynamics} or soft spots \cite{manning2011softspot}, induce long-range quadrupolar displacement fields of the type associated with Eshelby inclusions \cite{eshelby1957determination}, that provide equally many positive and negative couplings with other locations of plastic rearrangements.
% MM: I replaced this with:
Indeed, in amorphous solids units of plastic deformation -- referred to as shear transformation zones \cite{argon1979plastic,falk1998dynamics} or soft spots -- \cite{manning2011softspot}, induce long-range quadrupolar displacement fields of the type associated with Eshelby inclusions \cite{eshelby1957determination,maloney2006amorphous}, that provide equally many positive and negative couplings with other locations of plastic rearrangements.
The no-passing property must be violated in these systems, and one therefore expects that return point memory should not hold either. 
Yet there are experimental as well as numerical findings that are highly reminiscent of return point memory \cite{keim2018return,keimetalRMP2019}. 
Understanding memory effects in amorphous solids appears thus to require a deeper knowledge of the organization of states and transitions among these than is presently available. We develop such insights by introducing a novel method that maps the deformation paths of amorphous systems to directed graphs.  As recently shown by one of us, RPM is a well-defined property of such graphs that is easily identified \cite{munganterzi2018}.
We construct such graphs from simulations of sheared amorphous solids. Surprisingly, despite the fact that the coupling is not strictly positive, which precludes no-passing, these  systems show remarkably accurate, if not perfect, return point memory along with a near-perfect hierarchy of cycles and sub-cycles. 
% THIS IS THE SENTENCE TOM SUGGESTED:
We trace the smallest loops to local bistable hysteretic regions undergoing pure shear displacements \cite{manning2011softspot}.
%This can be explained by the fact that closed hysteresis loops, or limit cycles, are established by a collection of soft-spots \cite{manning2011softspot}, which act as two-level systems that switch between their states as the strain is varied and that in most cases the interactions do not cause RPM violations. 
% MM: I changed the sentence below as follows:
The relatively rare cases in which RPM is violated can be associated with certain destabilizing soft-spot interactions that lead to plastic events which provide 
%change their destabilizing strain values to such an extent that  provide 
one-way escapes from limit-cycles (``rabbit-holes'').

%\paragraph*{Simulations.} 
We simulate a two dimensional binary mixture of equal numbers of small and large particles (512 each) with size ratio 1.4, interacting with a radially symmetric potential (described in \cite{lerner2009locality,regev2013onset}). Energy minimum structures obtained from liquid configurations are subjected to  small strain steps of $\pm\Delta\gamma = 10^{-4}$ followed by energy minimization, implementing the athermal quasi-static (AQS) protocol used in previous studies. We thus always consider configurations at mechanical equilibrium at any given strain. 
%% MM August 28 this material is stated in the supplement
%\footnote{We verified that the transition graph is only minimally affected by the choice of minimization algorithm and strain step by comparing the results obtained from the FIRE and conjugate gradients algorithms and by repeating the simulations with $\Delta \gamma = 10^{-5}$}
Starting with a configuration at some strain $\gamma$, upon increasing strain, the configuration will deform elastically until a critical strain $\gamma^+$ is reached where a plastic rearrangement of particles occurs. Likewise, starting from the same initial configuration, and decreasing the strain, the system undergoes elastic deformations until a critical strain $\gamma^- < \gamma$ is reached, when another plastic event occurs
\footnote{Numerical implementation details are given in Supplement Section S1.1.}.

\begin{figure}[t]
\begin{center}
\includegraphics[width=\columnwidth]{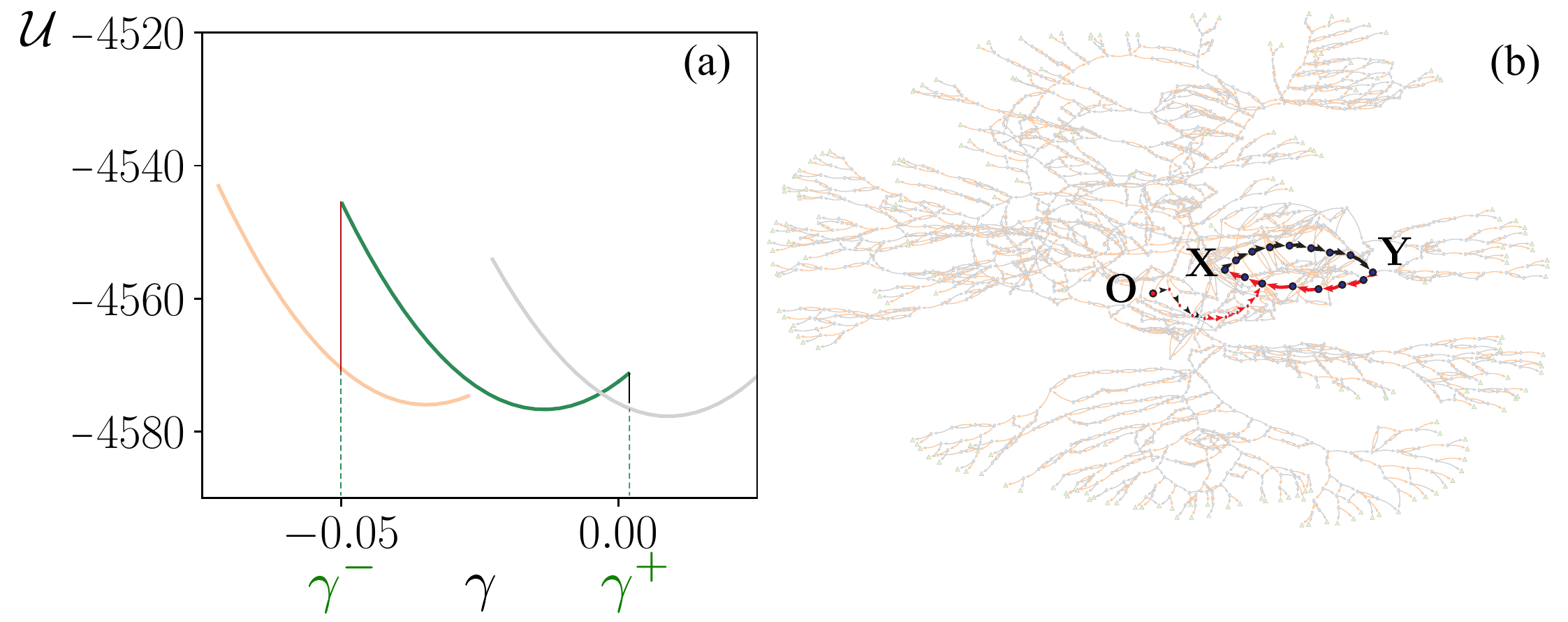} 
%\showthe\columnwidth % Use this to determine the width of the figure.
\caption{{\bf Mesostates and Transition Graphs}
(a) Potential energy $\mathcal{U}$ of particle configurations associated with an initial mesostate (green segment) and the two mesostates it transits into when the applied strain $\gamma$ ($x$-axis) becomes $\gamma^\pm$. 
(b) The network generated starting from a zero-strain configuration $O$. The  $1$ cycle transient and limit-cycle (between $X$ and $Y$) for $\gamma=0.05$ are marked in black and red.
}
\label{Fig1}
\end{center}
\end{figure}

%\paragraph*{Network construction.} 
We regard the set of stable configurations, whose members are continuously transformable into each other under strain changes, as one abstract state which we call a {\em mesostate}.
%\footnote{We shall use lower case letters to refer to microscopic configurations and label mesostates by upper case letters,}. 
The strain interval $\gamma^- < \gamma < \gamma^+$ over which purely elastic deformations are possible we call the stability range $\gamma^\pm$ of a mesostate. When a configuration of the mesostate is sheared to $\gamma^+$, a plastic event leads to a configuration, which belongs to a new mesostate. Likewise, straining in the negative direction to $\gamma^-$ leads to a plastic transition to a configuration belonging to a third mesostate. The potential energy associated with mesostates and their transitions is sketched in Fig~\ref{Fig1}(a). 
The mesostate transitions are {\em history-independent}: whenever the system is in a configuration $a$ belonging to mesostate $A$, it must transit to the same pair of mesostates when the strain is increased to $\gamma^+(A)$, or reduced to $\gamma^-(A)$.  These transitions can be represented as a graph where each node is a mesostate $A$ and two outgoing arrows specify the transitions to the mesostates that are reached after the plastic
events at $\gamma^{\pm}(A)$ \cite{munganterzi2018, munganwitten2018}. These transitions, together with their thresholds $\gamma^\pm$ suffice to prescribe the AQS response of the system to arbitrary shearing protocols \cite{munganterzi2018}.  
We use the numerical simulations to assemble a catalog of mesostates. 
For each state $A$ we record the values of $\gamma^\pm(A)$ and specify the two mesostates into which $A$ is mapped when $\gamma = \gamma^\pm(A)$. 
We limit our catalog to mesostates that can be reached from a chosen reference state $O$ in at most $\ell = 25$ transitions and
%These catalogs are then used to construct the state transition graph between mesostates. 
construct a transition graph from a reference configuration $O$ at zero-strain. A sample graph with $\mathcal{N} = 1416$ mesostates is shown in Fig. \ref{Fig1} (b) and exhibits tree-like features  as well as regions with high interconnections \footnote{See Supplement Section S2.1 for a blown-up version of Fig.~\ref{Fig1}(b) along with a comparison of the transition graph obtained from a system with a larger number of particles. }. A detailed discussion of general features will be done elsewhere \cite{longpaper2019amorph}.
%In addition, there are also bottlenecks, where transitions from one region of the graph to another have to proceed through a small number of intermediate states. 
Transitions under forward (positive) and backward (negative) shear are denoted by gray or orange arrows respectively. 
Certain transitions are emphasized by black and red highlights, respectively \footnote{We use green triangles to mark transitions to states not of interest, such as those beyond the catalog limit. 
See Supplement Section S1 for further details on the numerical procedure used for mesostate extraction, identification and transition graph construction.
}. 

%%%%%%%%%%%%%%%%%%%%%%%%%%%%%%%%%%%%%%%%%%%%%%%%
%%%%%%%%%%%%%%%%%%%%%%%%%%%%%%%%%%%%%%%%%%%%%%%%

%%%%%%%%%%%%%%%%%%%%%%%%%%%%%%%%%%%%%%%%%%%%%%%%
\begin{figure*}[t!]
\begin{center}
%\showthe\columnwidth % Use this to determine the width of the figure.
\includegraphics[width=2\columnwidth]{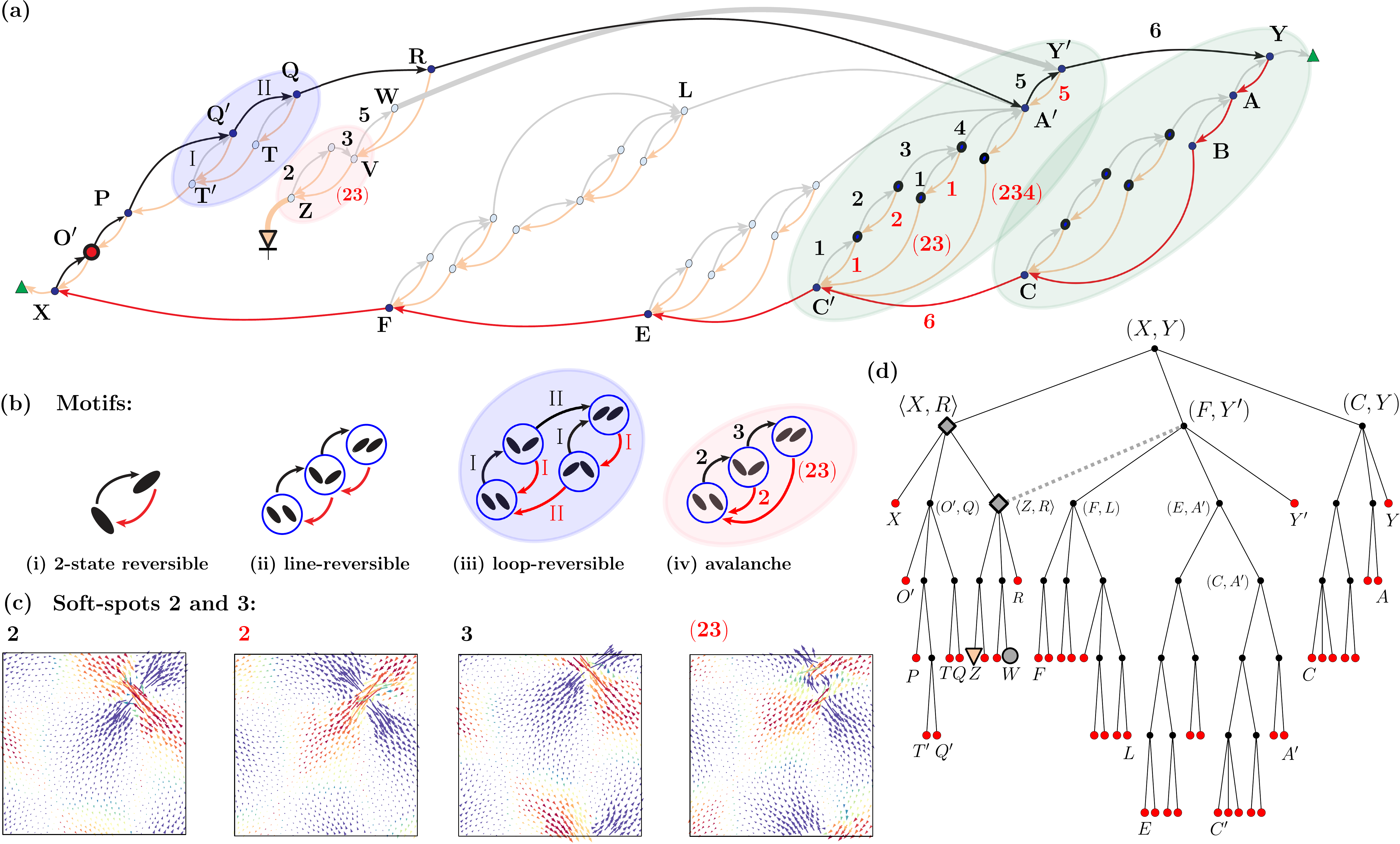}
\caption{
{\bf State transition graph of the $\gamma_{\rm max} = 0.05$ limit-cycle.}
(a) Detailed view of the mesostate transitions associated with the $0.05$ limit-cycles depicted in Fig.~\ref{Fig1} (b). Transitions out of the endpoints $X$ and $Y$ are marked as green triangles and will be ignored. Regions of interest have colored backgrounds and refer to (b) and Fig.~\ref{echos}(c).
(b) Network motifs involving one and two soft-spots, (i) and (ii) - (iv), respectively. Soft-spots are shown as black ellipses with states corresponding to their orientation. Motif background color and transition pattern highlighted in (a) coincide.
(c) The particle displacements associated with the transitions of the avalanche motif in (iv) and (a). 
(d) Tree representation of the hierarchy of loops and sub-loops making up the limit cycle shown in (a). Refer to text for details. 
} 
\label{lc050}
\end{center}
\end{figure*}
%%%%%%%%%%%%%%%%%%%%%%%%%%%%%%%%%%%%%%%%%%%%%%%%

%\paragraph*{Limit-cycle structure.}

To understand a typical limit-cycle in terms of the transition graph, starting in $O$ and using our catalog we can trace out the set of mesostates obtained for periodic shear with strain $0 \to \gamma_{\rm max} \to -\gamma_{\rm max} \to 0 \to \cdots $. 
Fig~\ref{lc050}(a) shows the mesostates and transitions of the $\gamma_{\rm max} = 0.05$ limit cycle and its vicinity.
%depicted in Fig.~\ref{Fig1} (b). 
With the limit-cycle in state $X$ at $ -\gamma_{\rm max}$ and as the strain increases, the system undergoes a sequence of plastic events (black arrows), passing through the mesostates $O',P, Q',Q, R, A', Y'$ and reaching the upper endpoint $Y$ at $ +\gamma_{\rm max}$. 
Subsequently reducing the strain back to $-\gamma_{\rm max}$, the dynamics follows the red arrows, passing through  $A, B, C, C', E, F$ and eventually returning to the lower endpoint $X$. Reversing the shearing direction anywhere along the decreasing (red) branch will lead to the upper end point $Y$. Likewise, reversal along the increasing (black) branch leads to $X$,  except for $R$, where the strain reversal leads via $Z$ to an exit from the loop. 
Trajectories that return to an endpoint upon a strain reversal necessarily form {\em sub-cycles}. For example the pair of states $(C,Y)$ are the endpoint of a sub-cycle.
%starting at $C$ and increasing the strain, we reach $Y$ by following the gray arrows. 
%A subsequent strain reduction brings the system back to $C$. 
In fact, a hierarchical structure of cycles nested within cycles is apparent. This structure is highly reminiscent of return point memory, as discussed below.
%%%%%%%%%%%%%%%%%%%%%%%%%%%%%%%%%%%%%%%%%%%%%%%%
\begin{figure}[t]
\begin{center}
%\showthe\columnwidth % Use this to determine the width of the figure.
\includegraphics[width=\columnwidth]{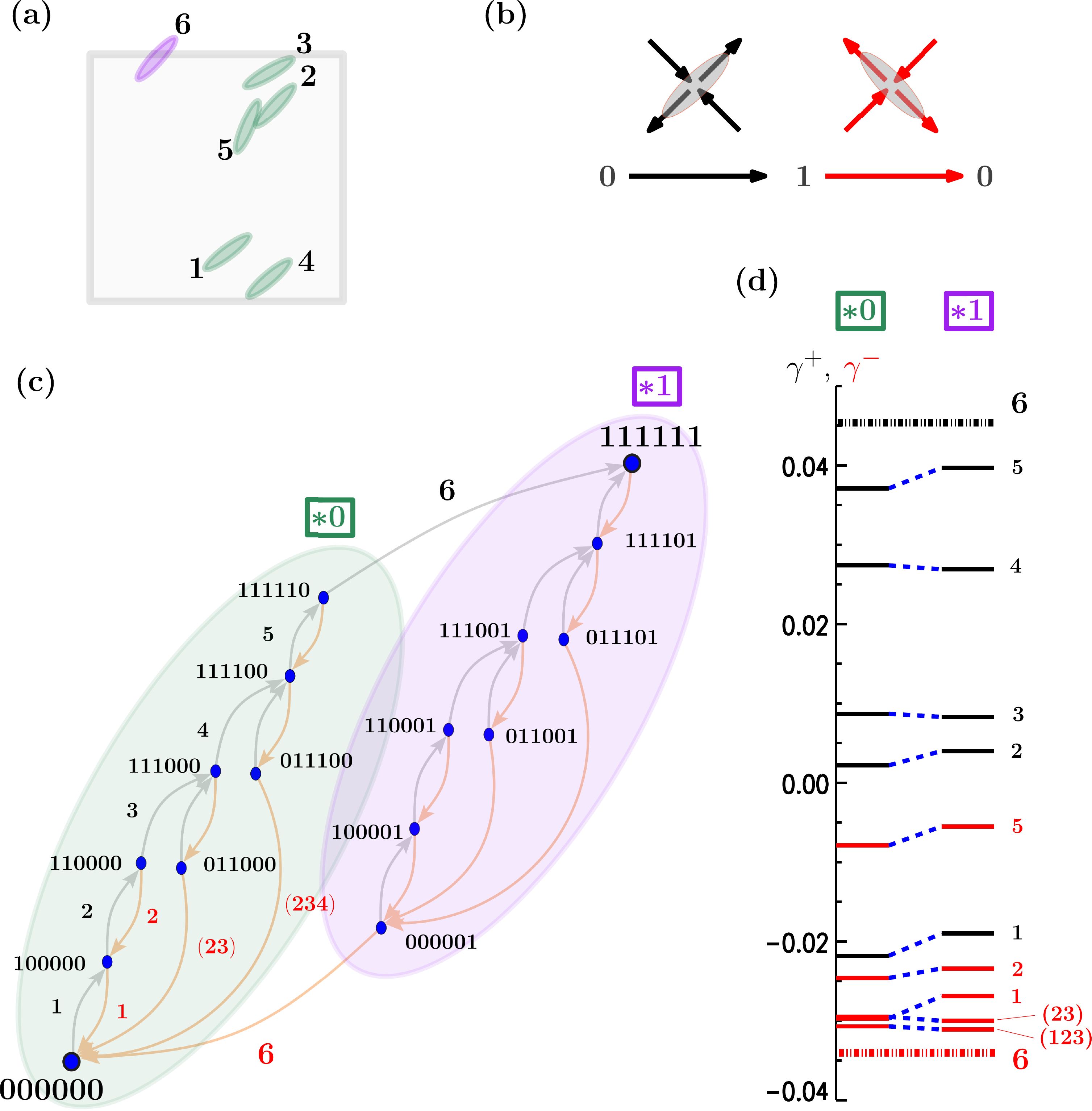}
\caption{
{\bf Limit-cycles as interacting soft-spot systems.} 
(a) Location and label of the six soft spots in the sample that produce the plastic events of the cycle $(C',Y)$ in Fig.~\ref{lc050}(a). 
(b) Schematic description of particle displacements during a state change of a soft spot  $0\rightarrow 1$ and back, $1\rightarrow 0$.
(c) State transitions in the sub-cycles marked $\bf{[*0]}$ and $\bf{[*1]}$.
%, shaded green and purple. (detail from Fig.~\ref{lc050}(a)). 
The two cycles are topologically identical and the transitions in each are due to the same 5 soft spots.  
Transition from one cycle to the other occurs via the state change of a sixth soft spot. 
Labels next to the transitions in sub-cycle $\bf{[*0]}$ indicate the soft spots involved (same in  sub-cycle $\bf{[*1]}$). 
Binary strings next to each mesostate indicate the individual states of soft-spots. 
The transitions marked $(23)$ and $(234)$ are avalanches.   
(d) ``Spectroscopic'' plot showing the non-monotonic changes in the switching strains $\gamma^{\pm}$ of soft-spot $1 - 5$, depending on the state of $\#6$. 
The horizontal lines indicate the values of the switching strains 
along with the soft spots involved.
} 
\label{echos}
\end{center}
\end{figure}
%%%%%%%%%%%%%%%%%%%%%%%%%%%%%%%%%%%%%%%%%%%%%%%%

%\paragraph*{Motifs and soft spots.}
The state transition graph of Fig.~\ref{lc050}(a) has several recurring network transition patterns or ``motifs'',  which we depict in panels (i) - (iv) of Fig~\ref{lc050}(b). Inspecting the corresponding particle displacements, 
%% MM May 20: I did not understand why we are referring to Fig. 2(a) below. 
%Fig.~\ref{lc050}(a), 
we see these motifs arising from transitions, with hysteresis, between two states in localised soft spots.  
The simplest motif is a reversible transition which involves only one soft spot, Fig.~\ref{lc050}(b)(i), such as transitions between states $X$ and $O'$ or $O'$ and $P$ in Fig~\ref{lc050}(a). Next, transitions between $X$ and $P$ turn out to involve two soft-spots that change states one after the other as the strain is increased, or subsequently reduced, leading to a  {\em line reversible} motif, depicted in Fig~\ref{lc050}(b)(ii). 
Another pattern involves two soft-spots which change their states in the same order during an increase or decrease of strain, leading to the {\em loop reversible} motif, Fig~\ref{lc050}(b)(iii), {\em e.g.} the pattern highlighted in blue in Fig~\ref{lc050}(a) involving  transitions between $T',Q',Q,$ and $T$. The last, and perhaps most important motif we observe is due to avalanches. Here two or more two-level systems change states one after the other in one direction of strain, but return together to their initial state upon strain reversal, see Fig~\ref{lc050}(b)(iv). 
The region highlighted in pink in Fig~\ref{lc050}(a)) as well as the transitions of Fig~\ref{lc050}(a) marked $(23)$ and $(234)$ in the sub-cycle $(C',Y')$ are avalanches. 
%\paragraph*{Interactions of soft spots.}
The presence of avalanches implies that soft spots interact with each other.
The state of one soft spot can enable or even disable the ability of another soft spot to switch states.  The interactions between soft spots are mediated via an Eshelby-like quadrupolar deformation fields, arising from a change of state of one soft spot during a plastic event. 
They are shown in Fig~\ref{lc050}(c) for the four transitions making up the avalanche motif of Fig~\ref{lc050}(b)(iv).
Here the soft-spot labeled 2 can switch its state back and forth when soft spot 3 remains in one state. 
However after 3 also changes its state, both 2 and 3 reverse their states together. 
Another, rather striking, example for such soft-spot interactions is the pair of loops $(C',Y')$ and $(C,Y)$ in Fig.~\ref{lc050}(a), shaded in green, which have  identical topology. 
Transitions in these loops are due to the same $5$ soft spots, including $\#2$ and $\#3$ of 
Fig.~\ref{lc050}(c). 
%, two of which (2 and 3) we have encountered before. 
We have labeled them accordingly as $1$--$5$ and 
marked the transitions they generate in loop $(C',Y')$. 
The change from one loop to the other is due to a sixth soft spot. 
The locations in the sample of all six soft spots are marked in Fig.~\ref{echos}(a). 
Panels (b) - (c) of Fig.~\ref{echos} illustrate the binary encoding of the mesostates in terms of the states of each soft spot (refer to caption for further details). Fig.~\ref{echos} (d) depicts the non-monotonic changes of switching strains $\gamma^\pm_i$ for soft spots $1 -5 $ in dependence on the state soft-spot $6$. This non-monotonic behavior is consistent with the quadrupolar nature of the elastic deformation (note that this is a more complicated example of loop-reversible dynamics).

%\paragraph*{Return-point-memory and hierarchy of loops.}

The property of the system to return to the cycle's endpoints upon reversal of the forcing, when starting from a mesostate on a limit-cycle or on any of its sub-cycles, is called {\em loop return point memory} $(\ell$RPM).  
It is a generalization of RPM that does not require the existence of the no-passing property \cite{munganterzi2018}. 
The limit-cycle $(X,Y)$ of Fig~\ref{lc050}(a) would have $\ell$RPM, 
if two transitions were rewired: the orange arrow from $Z$ should point to $X$, while  the gray arrow from $W$ should lead to $R$. The first rewiring ensures that a strain reversal at $R$ leads to the lower endpoint $X$, while the second rewiring makes sure that in any sub-cycle of the now corrected cycle $(R,X)$ strain reversals 
lead to its endpoints $X$ and $R$.  With 2 RPM violating transitions out of 84, the  limit-cycle of Fig~\ref{lc050}(a) exhibits near perfect RPM  
In fact, we have observed near-perfect RPM in limit cycles for strain amplitudes up to at least $\gamma_{\rm max} = 0.0722$; see Section S2.2 in Supplemental Material for an example.
%\footnote{We have observed near-perfect RPM in many other limit cycles and for strain amplitudes up to at least $\gamma_{\rm max} = 0.0722$. See Supplement S2.2 for an example.}. 
In Fig~\ref{lc050} (d) we display the tree representation of the loop hierarchy introduced in \cite{munganterzi2018} for the $\gamma_{\rm max} = 0.05$ limit-cycle whose endpoints are $(X,Y)$. Nodes of this
tree represent cycles. Starting from the root $(X,Y)$,
%of the tree, 
each generation represents a partition of the parent cycles into two or three sub-cycles. 
The tree thus depicts the hierarchy of cycles under this partition. 
In a network with perfect return point memory, this will be a strict genealogical tree with each child loop having precisely one parent loop.
The RPM violations alter this structure. We have indicated the loops involved in RPM violations by gray diamonds, placing their would-be endpoints in angular brackets. 

\begin{figure}[h]
\begin{center}
%\showthe\columnwidth % Use this to determine the width of the figure.
\includegraphics[width=\columnwidth]{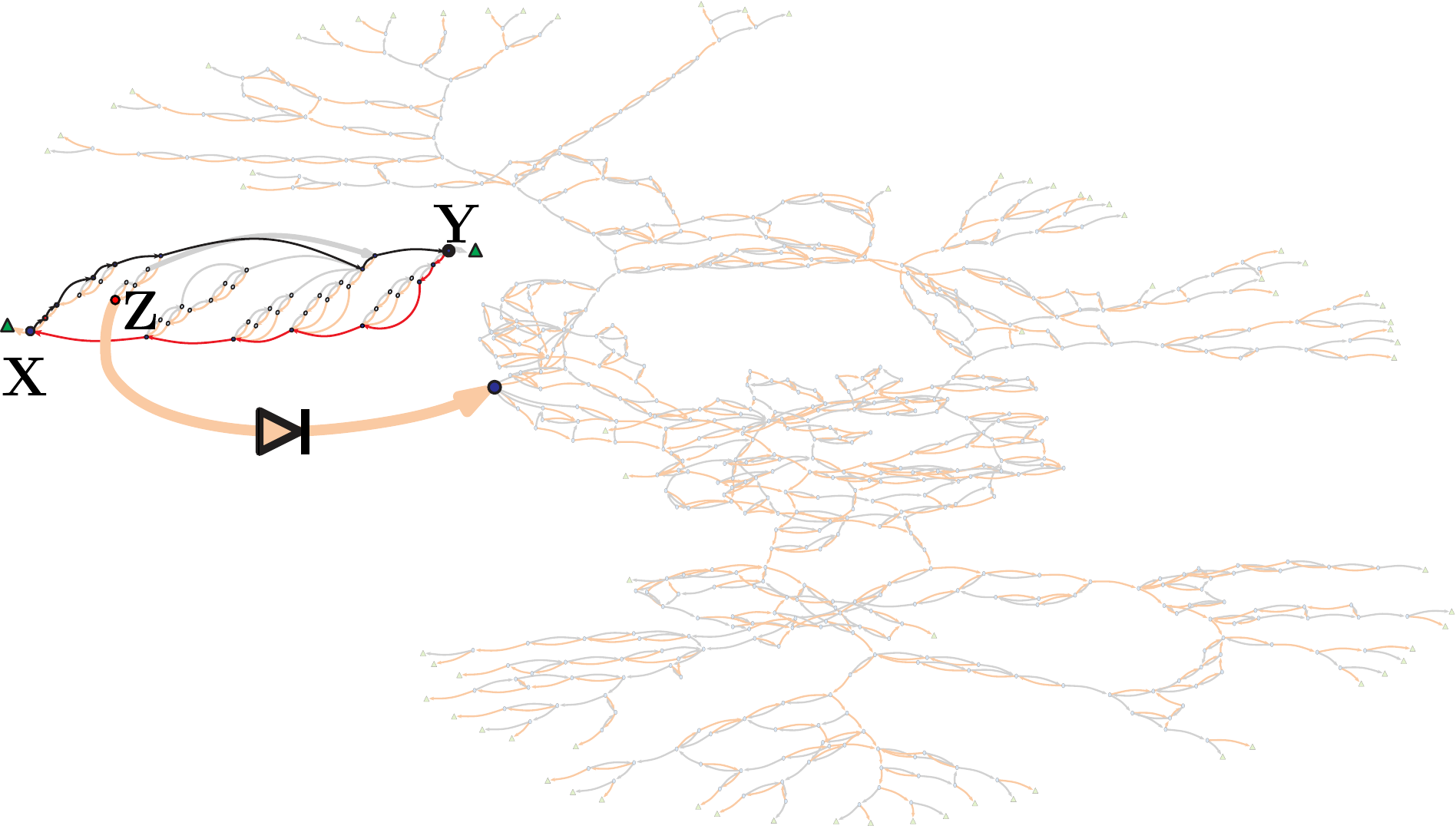}
\caption{
{\bf One-way ``rabbit-hole" transitions.} The transition out of the mesostate $Z$ inside the 
$\gamma_{\rm max} = 0.05$ limit-cycle leads to a set of states, from which there does not appear to be a path back.
%{\bf needs to be redone}
} 
\label{rabbit-hole}
\end{center}
\end{figure}

%\paragraph*{Violations of return-point-memory.}
The transition from $Z$ is like a step down a ``rabbit hole", as it leads to a part of the mesostate network from where (at $\ell = 25$) there seems to be no sequence of transitions that will bring the system back, {\em see} Fig.~\ref{rabbit-hole}.
This is a one-way transition and we have marked it with the diode sign. Such transitions have been discussed by Newman and Stein who noted that  multidimensional ragged landscapes involve one-way ``outlets'' \cite{stein1995broken}.
We believe that this one-way transition is caused by the 
permanent state change of one or more soft-spots. 
%%%%%%%%%%%%%%%%%%%%%%%%%%%%%%%%%%%%%%%%%%%%%%%%
%%%%%%%%%%%%%%%%%%%%%%%%%%%%%%%%%%%%%%%%%%%%%%%%
The transition from $W$ to $Y'$ is RPM-violating, since it takes a short-cut by going directly to $Y'$ without having $R$ as an intermediate
mesostate.
It turns out that the transition to $Y'$ via $R \to A' \to Y'$ involve the soft-spot labeled as $5$, {\em see} Fig \ref{lc050}(a), and another two soft-spots. 
However, the alternative route $R \to V \to W \to Y'$ is found to change the state of soft-spot $5$ during transition $V \to W$ under increasing strain. 
$W$ cannot transit into $R$ under a subsequent strain increase, because 
% MM: I am shortening this sentence. 
%the former differs from the latter by the state of soft-spot $5$ which has already changed its state when $V \to W$.  
soft-spot $5$ has already changed its state when $V \to W$.  
%\paragraph*{Discussion.} 

The construction of a network of mesostates that captures exclusively the plastic events,  provides a methodology that enables us to analyze in detail how memory is encoded in periodically sheared amorphous solids. We are thus able to explicitly address for the first time many questions about memory formation phenomena observed in these systems. In particular, we see that the system reaching a limit-cycle, is a result of the two-level nature of most plastic events and the property that in most cases, interactions modify the dynamics (and hence the network) in a manner that does not impair reversibility. As a result, a hierarchy of cycles and sub-cycles emerges.

At the same time, the transition network and its topology provide a bird's eye view on the athermal and quasistatic dynamics of an amorphous solid subject to arbitrary strain protocols. We believe that the prospect of relating the network topology to the activation and deactivation dynamics of interacting two-level systems provides a promising direction for understanding the reversible and irreversible features of the dynamics of amorphous solids and for constructing models that capture it.

\paragraph*{Acknowledgements.} The authors thank  Monoj Adhikari, Nathan Keim, Sid Nagel, Jim Langer, Mert Terzi, and Tom Witten  for comments and discussions. 
The authors acknowledge  KITP for its hospitality and generous support through grant  NSF PHY 17-48958.  
MM thanks the International Centre for Theoretical Sciences (ICTS) for supporting a visit and participation in the program - Entropy, Information and Order in Soft Matter ICTS/eiosm2018/08. 
MM was supported by the German Research Foundation (DFG) under DFG Projects No. 398962893 and the DFG Collaborative Research Center 1060 ``The Mathematics of Emergent Effects".  IR was supported by the Israel Science Foundation (ISF) through Grant No. 1301/17 and the German-Israel foundation (GIF) through Grant no.I-2485-303.14/2017. KD thanks the US National Science Foundation for support through grant NSF CBET 1336634.
SS acknowledges support through the JC Bose Fellowship DST (India).

\paragraph*{Author contributions.}
MM and IR contributed equally to the research of this work. KD, MM, IR and SS analyzed the results and wrote the paper.

\bibliography{AmorphNets}

\end{document}